\documentclass[12pt,preprint]{aastex}

\shorttitle{vector modes and CMB multipoles}
\shortauthors{J.A. Morales and D. S\'aez}

\begin{document}


\title{Large scale vector modes and the first\\ 
CMB temperature multipoles}


\author{J.A. Morales and D. S\'aez}
\affil{Departamento de Astronom\'{\i}a y Astrof\'{\i}sica, Universidad de Valencia,
46100, Burjassot, Valencia, Spain}
\email{diego.saez@uv.es}  
\email{antonio.morales@uv.es}  



\begin{abstract}

Recent observations have pointed out various anomalies 
in some multipoles (small $\ell $)   
of the cosmic microwave background (CMB). 
In this paper, it is proved that some of these 
anomalies could be explained in the framework of
a modified concordance model, in which, there is an appropriate
distribution of 
vector perturbations with very large
spatial scales.
Vector modes are associated with 
divergenceless (vortical) velocity fields. 
Here, the generation of these modes is not studied in 
detail (it can be done ``a posteriori"); on the contrary,
we directly look for the distributions of these vector modes which     
lead to both alignments of the second and third 
multipoles and a planar octopole. 
A general three-dimensional (3D) superimposition of vector perturbations 
does not produce any alignment, but we have found 
rather general 2D superimpositions
leading to anomalies similar to the observed ones; in these 2D cases,    
the angular velocity has the same direction 
at any point of an extended region and, moreover, 
this velocity has the same distribution in all the 
planes orthogonal to it. 
Differential rotations can be seen as particular 
cases, in which, the angular velocity only depends on the distance to 
a rotation axis. Our results strongly suggest that appropriate mixtures of 
scalar
and vector modes with very large spatial scales could explain the 
observed CMB anomalies.

\end{abstract}





\section{Introduction}
\label{sec1}

The analysis of the data obtained 
by the Wilkinson Microwave Anisotropy Probe (WMAP) 
has
pointed out some anomalies in the temperature distribution
of the Cosmic Microwave Background (CMB). These anomalies 
have not been explained
in the framework of the concordance model, which is an inflationary 
flat universe 
with cold dark matter, dark energy, and reionization. 
For appropriate values of the involved parameters, this 
model explains most of the current cosmological observations, e.g., 
the magnitude-redshift relation satisfied by far
supernovae, 
the statistical properties of galaxy surveys, 
and the CMB anisotropies; nevertheless, some aspects of these observations 
remain controversial. Among them,
the WMAP anomalies deserve attention. Some of these anomalies could be 
due to unexpected systematic errors associated to 
foreground subtraction, galactic cuts, statistical analysis, and so on; 
however, other anomalies 
could be true effects requiring new physics.
Future experiments as
PLANCK should distinguish between physical effects
and systematic errors. Let us now list the main anomalies:
(i) the amplitude of the $C_{2} $ multipole 
is lower than it was expected, (ii) 
there is an asymmetry between 
the North and South ecliptic hemispheres,
(iii) the multipole
$C_{3} $ is too planar, and (iv) the multipoles
$C_{2} $ and $C_{3} $ are too aligned. Other anomalies 
concerning $\ell > 3$ multipoles have been also described.

The importance of the anomaly (i) was initially overestimated. 
The probability assigned by \citet{spe03} to the $C_{2} $ value obtained 
from the first year WMAP data was $\sim 1.5 \times 10^{-3}$. Afterward, 
other authors \citep{efs03,gaz03,efs04,slo04} obtained greater probabilities 
by using different methods for data analysis. Finally, \citet{hin06} used 
the data from the first three years of the WMAP sky survey, 
plus appropriate statistical and foreground subtraction techniques, 
to conclude that 
the probability of the measured $C_{2} $ multipole is 
$\sim 0.16 $. In conclusion, the observed value of $C_{2} $
is currently considered small but compatible with 
the concordance model. Nevertheless, 
a lack of correlations at 
the largest angular scales appears to be statistically significant in cut-sky 
maps (see \citet{spe03,cop07,haj07})

The anomaly (ii) was studied in detail by \citet{eri04,eric04,han04,hans04}. 
The hemispherical power asymmetry is nowadays considered 
substantial and robust, nevertheless, more study is necessary to
get definitive conclusions \citep{eri07}. 

Mathematical methods to quantify the alignment of
$C_{2} $ and $C_{3} $ as well as the planar character of $C_{3} $
were depicted by \citet{oli04} [vectors $\vec{n}_{2}$, $\vec{n}_{3}$
and parameter $t$] and \citet{cop04} [multipole vectors]. 
For the sake of simplicity, we have designed a code to
compute $\vec{n}_{2}$, $\vec{n}_{3}$ and $t$, whereas multipole vectors
will be considered elsewhere. Vectors $\vec{n}_{2}$ and $\vec{n}_{3}$
make maximum  quantity 
\begin{equation}
\Psi = \sum_{m} m^{2} |a_{\ell m} (\vec{n})|^{2} 
\label{Ka}
\end{equation}
for $\ell = 2$ and $\ell = 3$, respectively.
In this last equation, quantities $a_{\ell m}(\vec{n})$ are the spherical
harmonic coefficients of the CMB map in a coordinate system 
where $\vec{n}$ coincides with the $z$-axis. See \citet{oli04} for the
explicit definition of parameter $t$. 
Anomalies (iii) and (iv) have been studied in 
many papers \citep{sch04,bie04,cop06,cop07}.
The planar shape of $C_{3} $ has been confirmed in the 
bibliography, but this characteristic of the octopole 
is not very unlikely in the concordance model. More 
problematic is the strong alignment of $C_{2} $ and 
$C_{3} $. Some authors state that the multipole alignment
is actually anomalous and also that the alignment extends up
to $\ell = 5$. They suggest the existence of a symmetry axis
\citep{lan05,lan06,ber07,cho07}. Other authors \citep{rac07} 
propose the existence of a preferred plane without rotational
symmetry. This proposal suggests either a differential rotation 
viewed from an arbitrary point of
the space, which should be outside the rotation axis, or 
a more complicated vortical motion with aligned angular
velocities. Motions of this 
type --in extended regions-- can be simulated with appropriate 
combinations of large scale 
vector modes. 

Finally, let us mention another CMB anomaly which has
been found at smaller angular scales: a non Gaussian 
cold spot ($\sim 10^{\circ}$ size) located in the South hemisphere 
\citep{vie04,cru05,mar06}.

An anisotropic Bianchi $VII_{h}$ 
model has been recently considered \citep{jaf05,bri05,jaf06,jaff06,gho07}
with the essential aim of explaining most of the 
above WMAP anomalies;
however, the authors recognize that their model does not 
explain the observed acoustic peaks.
Other authors have studied the anisotropy produced by big voids
with appropriate locations \citep{ino06,inou06} to account for 
the mentioned anomalies. Motivated by the above
considerations about symmetries and vortical (divergenceless) motions, 
we propose here
another possibility which may contribute to explain the
large angular scale CMB structure: the existence of vector 
perturbations with large enough spatial scales. Here,
the main features of the first $\ell $ multipoles 
produced by these vector modes are estimated 
in the framework of a concordance model. 

By using appropriate large scales, only their 
contribution to the first multipoles are 
significant and, consequently, there are no problems with 
the acoustic peaks. In the linear regime,
scalar, vector and tensor modes \citep{bar80} do not couple
among them; hence, vector modes can be separately studied.
Vector modes are vortical peculiar velocity fields which 
do not appear in standard inflation; nevertheless, 
large scale vector modes may appear in
brane-world cosmologies \citep{maa00} and also in models with
appropriate topological defects \citep{bun02}.
Whatever the origin of the vector modes may be,
we are interested in their possible effects on the CMB
when their amplitudes are appropriately normalized. 
Two effects produced by the same type of large scale vector perturbations 
were studied in \citet{mor07}. Some basic aspects concerning 
these perturbations can be found in this reference 
(hereafter, paper I).

Our background is the so-called
{\em concordance cosmological model}, with 
a reduced Hubble constant
$h=10^{-2}H_{0}=0.71$ (where $H_{0}$ is the
Hubble constant in units of  $Km \ s^{-1} Mpc^{-1}$).
The density parameters of vacuum energy and matter (baryonic plus dark)
are $\Omega_{\Lambda} = 0.73$ and $\Omega_{m} = 0.27$, respectively.
All these parameters are compatible with the analysis of the three first year
WMAP data recently published \citep{spe06}.

Along this paper, Greek (Latin) indexes run from $0$ to $3$ ($1$ to $3$).
Units are defined in such a way that $c= \kappa = 1$ where $c$ is
the speed of light and $\kappa = 8 \pi G / c^4$ is the Einstein
constant. The unit of length is the Megaparsec. 
Symbols $a$, $\eta $, and $z$ stand for the scale 
factor, the conformal time, and the redshift, respectively. Whatever quantity
$A$ may be, $A_{0} $ ($A_{e}$) stands for the value of $A$ at present (CMB emission)
time. 
Quantity $a_{0} $ is assumed to be unity. This choice is always
possible in a flat background.

\section{CMB anisotropy}
\label{sec2}

The most general vector perturbation of a FRW universe
is a fluctuation of the metric $g_{\alpha \beta}$, the four-velocity
$u^{\alpha} $, and the traceless tensor 
$E_{\alpha \beta} $ describing anisotropic stresses.

In the absence of scalar and tensor perturbations, 
the gauge can be chosen in such a way
that the line element reduces to
\begin{equation}
\label{linel}
ds^{2} = a^{2}(-d \eta^{2} + 2h_{i} dx^{i} d\eta + \delta_{ij} dx^{i} dx^{j}) \ ,
\end{equation}
where the perturbations of the $g_{0i}$ metric components have been written in
the form $h_{i}=(h_1, h_2, h_3) = \vec{h}$.

From the matter four-velocity, $u^{\alpha }=(u^{0}, \vec{u})$, 
one defines the peculiar velocity
$\vec {v} = \vec{u} / u^{0}$. 

Finally, the condition $E_{ij}=0 $ is assumed along 
the paper, which means that there are no anisotropic stresses 
conditioning the evolution of $\vec{v} $ and $\vec{h} $.

Let us now calculate the CMB temperature contrast, $\Delta T/T$, due to
the above linear vector perturbations.    
From the equations of the null geodesics, the following formula
can be easily obtained:
\begin{equation}
\label{dtt}                                                                   
\frac {\Delta T}{T} =\vec {v}_{c0} \cdot \vec{n} - \vec {v}_{ce} \cdot \vec{n}
- n^{i} n^{j} \int^{\eta_{e}}_{\eta_{0}} \frac {\partial h_{j}}{\partial x^{i}} 
d \eta   \ ,
\end{equation}
where $\vec{n}$ is the unit vector in the observation direction 
and $\vec{v}_{c} = \vec{v} + \vec{h}$. In the case of linear vector modes,
the integral can be calculated along radial null geodesics
of the FRW background, whose equations are 
$ \dot{\eta } = - \dot{r}$, $\dot{\theta} = \dot{\phi} = 0 $ (in terms
of the spherical coordinates $r$, $\theta $ and $\phi $ associated
to $x^{i} $). The dots stand for derivatives with respect to the 
affine parameter.
                     
Functions $\vec{h}(\eta,\vec{r}) $ and $\vec{v}(\eta,\vec{r}) $ can be 
expanded in terms of an appropriate basis (the fundamental harmonic vectors,
see \citet{bar80} and \citet{hu97})
to write
\begin{equation}
\label{exph}
\vec{h}(\vec{r},\eta ) = - \int [ B^{\,+}(\vec{k},\eta) \vec{\epsilon}^{\,+}
(\vec{\kappa})+ B^{\,-}(\vec{k},\eta) \vec{\epsilon}^{\,-}
(\vec{\kappa})]
\exp (i \vec{k} \cdot \vec{r}) \, d^{3} k    \ ,                          
\end{equation}
where $\vec{k} $ is the wavenumber vector, $\vec{\kappa}$ is the unit
vector $\vec{k} /k$, and functions $B^{\,+}$ and $B^{\,-}$ are the coefficients
of the $\vec{h}$-expansion. A representation of vectors 
$\vec{\epsilon}^{\,+}$ and $\vec{\epsilon}^{\,-}$ is 
\begin{equation}
\epsilon^{\pm}_{1}=(\pm k_{1} k_{3} /k - i k_{2})/  \sigma \sqrt{2}
\label{ep1}     \ ,
\end{equation}
\begin{equation}
\epsilon^{\pm}_{2}=(\pm k_{2} k_{3} /k + i k_{1})/ \sigma \sqrt{2}
\label{ep2}        \ ,
\end{equation}
\begin{equation}
\epsilon^{\pm}_{3}=\mp  \sigma / k \sqrt{2} 
\label{ep3}     \ ,
\end{equation} 
where $\sigma = (k_{1}^{2}+k_{2}^{2})^{1/2}$ 
(see paper I). Hereafter, the following compact notation is
used $B^{\,+}(\vec{k},\eta) \vec{\epsilon}^{\,+}
(\vec{\kappa})+ B^{\,-}(\vec{k},\eta) \vec{\epsilon}^{\,-}
(\vec{\kappa})= B^{\,\pm} \vec{\epsilon}^{\,\pm}$.
Vector $\vec{v}(\eta,\vec{r}) $ is expanded in the same way using 
the coefficients $v^{\,\pm}(\vec{k},\eta)$. Quantities 
$v_{c}^{\,\pm} = v^{\,\pm} - B^{\,\pm}$
are gauge invariant \citep{bar80}. 
Under the condition $E_{ij} = 0$, quantities $B^{\,\pm} (\vec{k}, \eta)$
decrease as $a^{-2}$ in both the radiation dominated and 
the matter dominated eras (see paper I). 
Therefore,
vector metric perturbations being significant at decoupling (the end of inflation) 
would be negligible today (at decoupling).
During matter domination, the following formula holds:
$B^{\,\pm} (\vec{k}, \eta) = 6 H_{0}^{2} \Omega_{m}
v_{c0}^{\,\pm}(\vec{k})/k^{2}a^{2}(\eta)$.
Furthermore, functions $v_{c}^{\,\pm}$
are proportional to $a^{-1}$ (constant) in the matter (radiation)
dominated era. 
According to these comments,
vector modes producing significant effects on the CMB should not freely evolve 
from the early universe. Either they are produced 
by exotic processes (brane-worlds, strings, and so on) close enough to 
recombination-decoupling or they must be maintained by
some field producing an appropriate $E_{ij} \neq 0$ vector component
(see paper I).
Using the above expansions and evolution laws, the relative temperature
variation due to the last term of Eq. (\ref{dtt}) can
be rewritten as follows: 
\begin{equation}
\label{int3} \frac {\Delta T}{T} = 6 H_0^2 \Omega_{m} \, \int_{0}^{r_e}
\frac {dr} {a^{2}(r)} F(\vec{r})  ,
\end{equation} 
where $F(\vec{r}) = F_{pq}(\vec{r}) n^{p} n^{q}$ and
\begin{equation}
\label{int33}
F_{pq}(\vec{r}) = -i \int \frac {k_{p}} {k^{2}} \, v_{c0}^{\,\pm}(\vec{k}) \,
\epsilon^{\,\pm}_{q}(\vec{\kappa}) \, \exp (i \vec{k} \cdot \vec{r}) \, d^{3} k
\ .
\end{equation}
This last equation can be seen as a Fourier transform for each pair $(p,q)$ of indexes.
After these transforms are performed for appropriate boxes and resolutions,
function $F(\vec{r})$ and the integral in Eq. (\ref{int3}) can be 
easily calculated for a set of directions defining a sky CMB map.
A HEALPIx ({\em Hierarchical Equal Area Isolatitude
Pixelisation of the Sphere}, see \cite{gor99} ) pixelisation
covering the sky with 3072 pixels is used in our simulations.

Apart from the above CMB temperature effects, vector modes produce 
a rotation of the polarization direction (Skrotskii effect). 
As it was proved in paper I, the rotation angle is                    
\begin{equation}                            
\label{int6} \delta \psi = 3 H_0^2 \Omega_{m} \, \int_{0}^{r_e}
\frac {dr} {a^{2}(r)} [\vec{n} \cdot \vec{G}(\vec{r})]  ,
\end{equation} 
where
\begin{equation}
\label{int66} \vec{G}(\vec{r}) = \int
 \frac
{v_{c0}^{\, +} \, \vec{\epsilon}^{\,\, +}(\vec{\kappa})
-v_{c0}^{\, -} \, \vec{\epsilon}^{\,\, -}(\vec{\kappa})}{k}
 \exp (i \vec{k} \cdot
\vec{r}) \, d^3 k  .
\end{equation}

For the line element (\ref{linel}), 
the components of the angular velocity in momentum space are
$W_{i} = a^{3} \epsilon_{ijk} W^{jk} $. From this relation and
the $W_{mn} $ components given by  
\citet{bar80}, one easily gets --at first order-- the following formulas: 
\begin{equation}
W_{1}=iv_{c}^{\,\pm}(\epsilon_{2}^{\,\pm}k_{3}-\epsilon_{3}^{\,\pm}k_{2})
\label{w1}     \ ,
\end{equation}
\begin{equation}
W_{2}=iv_{c}^{\,\pm}(\epsilon_{3}^{\,\pm}k_{1}-\epsilon_{1}^{\,\pm}k_{3})
\label{w2}        \ ,
\end{equation}
\begin{equation}
W_{3}=iv_{c}^{\,\pm}(\epsilon_{1}^{\,\pm}k_{2}-\epsilon_{2}^{\,\pm}k_{1})
\label{w3}     \ ;
\end{equation} 
hence, the equation $k_{i} W_{i} =0$ is identically satisfied.
The resulting components only depend on the gauge invariant quantities
$v_{c}^{\,\pm} = v_{c0}^{\,\pm}(\vec{k})/a$ and, consequently, 
the angular velocity is an appropriate vector field in order
to discuss the properties of the vector modes and their superimpositions  
in a gauge invariant way (it is not the case of the peculiar 
velocity). 

Various appropriate
choices of $v_{c0}^{\,\pm}(\vec{k})$ are considered in next sections. In each case,
the angular velocity and the resulting $\Delta T/ T$ and $\delta \psi$ maps are analyzed. 
For the $\Delta T/ T$ maps, the angle formed by vectors $\vec{n}_{2}$ and $\vec{n}_{3}$
(giving the directions of the quadrupole and octopole)
and the parameter $t$ defining the planar character of the octopole (see \citet{oli04}) 
are calculated.

\section{CMB anisotropy produced by a single vector mode}
\label{sec3}

An unique vector mode $\vec{k}_{u}$ is first considered. In this way, 
some ideas --which are basic in next sections to understand the 
CMB effects produced by superimpositions of these modes-- 
are pointed out. For an unique mode, we can write:
\begin{equation}
\label{vdelta} v^{\pm}_{c u} (\vec{k})= v^{\pm}_{c u} \delta
(\vec{k} - \vec{k}_u) -(v^{\pm }_{c u})^{*} \delta (\vec{k} +
\vec{k}_u),
\end{equation}
where the complex numbers $v^{\pm}_{c u} = v^{\pm}_{cuR} + i
v^{\pm}_{cuI}$ fix the amplitude of the chosen mode and $\delta
(\vec{k} - \vec{k}_u)$  and $\delta (\vec{k} + \vec{k}_u)$ are
Dirac-distributions. Equation (\ref{vdelta}) implies the relation
[$v^{\pm }_{c u} (\vec{k} )]^{*} = - v^{\pm}_{c u} (- \vec{k} )$,
which ensures that the components of the angular velocity in 
position space, as well as
the temperature contrast $\Delta T/ T$ and the Skrotskii rotation angle 
$\delta \psi$ are real numbers. Moreover, for an unique mode, 
the coordinate axis in momentum space can be chosen in such a
way that $\vec{k}_{u} = (k_{u1}, 0, 0)$ with $k_{u1}=k_{u}>0$ and, then, Eqs.
({\ref{ep1})--({\ref{ep3}) leads to:
\begin{equation}
\epsilon^{\pm}_{1}=0; \,\,\,\, \epsilon^{\pm}_{2}=i/\sqrt{2}; \,\,\,\,
\epsilon^{\pm}_{3}=\mp 1/\sqrt{2} \ .
\label{simep}
\end{equation}
For the sake of simplicity in the notation,
the $x_{1}$, $x_{2}$, and $x_{3}$ components of the angular velocity  
are hereafter denoted $W_{x}$, $W_{y}$, and $W_{z}$, respectively.
The same notation is used for the components of any other 
vector in position space.
From Eqs. (\ref{w1})--(\ref{simep}) one easily gets:
\begin{equation}
W_{x} = 0
\label{wx} 
\end{equation} 
\begin{equation}
W_{y} = k_{u} \sqrt{2} \,\, \Big[ (v_{cuR}^{+}-v_{cuR}^{-})\sin{\xi}
+(v_{cuI}^{+}-v_{cuI}^{-})\cos{\xi}  \Big]
\label{wy} 
\end{equation}   
\begin{equation}
W_{z} = k_{u} \sqrt{2} \,\, \Big[ (v_{cuR}^{+}+v_{cuR}^{-})\cos{\xi}
-(v_{cuI}^{+}+v_{cuI}^{-})\sin{\xi}  \Big]
\label{wz} 
\end{equation}                
where $\xi = \vec{k}_{u} \cdot \vec{r} = k_{u} \, r \sin\theta \cos\phi $ 
and variables $r$, $\theta$, and $\phi$ 
are spherical coordinates in position space.
Analogously, From Eqs. (\ref{int33}), (\ref{vdelta}) and (\ref{simep})
one proves that the only non-vanishing components of $F_{pq}(\vec{r})$ are
$F_{12}=W_{z}/k_{u}^{2}$ and $F_{13}=-W_{y}/ k_{u}^{2}$.
As it follows from these relations and Eqs. (\ref{wy})--(\ref{wz}),
functions $F_{12} $ and $F_{13} $ depend on our choice of the complex
numbers $v_{cu}^{+} $ and $v_{cu}^{-} $. Once these numbers have been 
chosen, the integral of the r.h.s. of Eq. (\ref{int3}) can be easily
written as follows:
\begin{equation}
\frac {\Delta T}{T} = \frac {6 \sqrt{2} H_0^2 \Omega_{m} n^{1}} {k_{u}}
\Big[ (A\, n^{2} + B\, n^{3}) I_{c} + (C \, n^{2} + D\, n^{3}) I_{s} \Big] \ ,
\label{int4}
\end{equation}
where $A=v_{cuR}^{+} + v_{cuR}^{-}   $, 
$B= v_{cuI}^{-} - v_{cuI}^{+}  $, 
$C= -(v_{cuI}^{+} + v_{cuI}^{-})  $, 
$D= v_{cuR}^{-} - v_{cuR}^{+}  $,
\begin{equation}
\label{isin} I_{s} = \int_{0}^{r_e}  a^{-2}(r) \sin \xi \, dr \ ,
\end{equation}
\begin{equation}
\label{icos} I_{c} = \int_{0}^{r_e}  a^{-2}(r) \cos \xi \, dr \ ,
\end{equation} 
$n^{1} = \sin{\theta} \cos{\phi}$, $n^{2} = \sin{\theta} \sin{\phi}$, and
$n^{3}=\cos{\theta}$.
The integrals (\ref{isin}) and (\ref{icos}) are to be performed
along each of the 3072 directions configuring our HEALPIx map
from emission ($r_{e} $) to observation ($r = 0 $).
Afterward, the resulting map can be analyzed by using our numerical code 
specially designed to get 
$\vec{n}_{2}$, $\vec{n}_{3}$, and $t$.
    
The value $k_{u} = 2 \pi /L_{u} $ with $L_{u} = 4 \times 10^{4} \ Mpc $ has 
been fixed and, then,    
for $A=C=0$ and $B=D=6.6 \times 10^{-10}$ (mode [1]),
vectors $\vec{n}_{2}$ and $\vec{n}_{3}$ 
appear to be perfectly aligned in the direction (0,1,0) and the octopole
is rather planar ($t=0.93$). 
The total $\Delta T/T$ map is displayed in the top panel
of Fig.~\ref{f1}. The central and bottom panels of the same figure show the 
quadrupolar and octopolar components of this map.
Figure \ref{f2} has the same structure but it corresponds 
to $A=B=-C=D=3.3 \times 10^{-10}$ (mode [2]). In this last case,
there is no alignment. The angle formed by the vectors 
$\vec{n}_{2} = (0.037, 0.706, 0.707)$ and 
$\vec{n}_{3} = (-0.037, 0.706, -0.707)$ is very close to
$90^{\circ}$ and parameter $t$ takes on the value $t=0.93 $
(as in the first case). Other angles and $t$ values appear
for other choices of parameters $A,B,C$ and $D$.
These results strongly suggest that random superimpositions of 
arbitrary vector modes 
should not lead to aligned $\vec{n}_{2}$ and $\vec{n}_{3}$ vectors. 
This fact is verified in next section by considering a rather general 3D
superimposition.

Finally, another type of vector modes 
(hereafter called $w$-modes) deserves particular 
attention (see \S ~\ref{sec5} for applications). 
In this case, the  
coordinate axis in momentum space are chosen in such a way that
$\vec{k}_{u} = (k_{u1}, k_{u2}, 0)$ and, then, the conditions
$v_{cu}^{+}  = v_{cu}^{-} = v_{cu}$ are assumed. Complex number
$v_{cu}$ can be put in the form $v_{cu}=|v_{cu}| \cos{\beta}
+ i \, |v_{cu}| \sin{\beta} $. Similarly, we can write 
$k_{u1} = \sigma_{u} \cos{\alpha} $ and $k_{u2} = \sigma_{u} \sin{\alpha} $.
The effect of an unique $w$-mode is now considered.
By performing the same kind of calculations as 
for previous isolated modes, one easily get:
\begin{equation}
W_{x} = W_{y} = 0
\label{wxy} 
\end{equation} 
\begin{equation}
W_{z} = 2 \sqrt{2} \sigma_{u} |v_{cu}| \,\, (\cos{\beta} \cos{\xi} 
- \sin{\beta} \sin{\xi}) \ ,
\label{wz2} 
\end{equation}    
where $\xi = \vec{k}_{u} \cdot \vec{r} =  \, \sigma_{u} \, r \, (n^{1} \cos{\alpha} + n^{2} \sin{\alpha} )$.
Furthermore, the associated temperature contrast is:
\begin{equation}
\frac {\Delta T}{T} = \frac {12 \sqrt{2} H_0^2 \Omega_{m} |v_{cu}|} {\sigma_{u}}
\Big[I_{c} \cos{\beta}  - I_{s} \sin{\beta} \Big] 
\Big[  (n^{2} n^{2} -n^{1} n^{1}) \frac {\sin{2 \alpha}}{2}+
 n^{1} n^{2} \cos{2 \alpha}\,  \Big] \ .
\label{int5}
\end{equation}
Thousands of maps, $M_{i}$, corresponding to different values 
of $\alpha $ and $\beta $
have been obtained and analyzed. Parameters $\sigma_{u}$
and $|v_{cu}|$ have been fixed. Their values are 
$\sigma_{u}=\pi /(2 \times 10^{4}) \ Mpc^{-1}$ and 
$|v_{cu}| = 3.3 \times 10^{-10}$.
In Fig.~\ref{f3} we display three of 
these maps corresponding to distinct $w$-modes; they are different, 
but the spots are always aligned along 
the equatorial zone and, consequently, 
as it has been verified, 
vectors $\vec{n}_{2}$ 
and $\vec{n}_{3}$ are aligned along the direction $(0,0,1)$ and,
moreover, the octopole is very planar $t \simeq 0.94 $.
This type of alignment and a high $t$ value (planar octopole) 
appear in all the maps. Other values of $\sigma_{u}$
and $|v_{cu}|$ have been considered with the same result.
If we superimpose many of these maps, the vectors 
$\vec{n}_{2}$ and $\vec{n}_{3}$ of the resulting map are not always 
aligned; in other words, any combination of linear modes lying in
the plane $(k_{1},k_{2})$ with $v_{c0}^{+}  = v_{c0}^{-} = v_{c0}$ 
does not lead to multipole alignments.

This fact is not surprising taking into account
that, for a given map, directions $\vec{n}_{2}$ and $\vec{n}_{3}$ maximize
the quantity $\Psi $ defined in Eq. (\ref{Ka}), which is {\em nonlinear} with
respect to the $a_{\ell m}$ coefficients. 
Superimpositions of $w$-modes have been numerically analyzed 
in a simple way, we have 
taken 1521 maps $M_{i}$ and, then, other 1521 maps $N_{j}$ have 
been obtained according to the following formula: 
$N_{j} = \sum_{i=1}^{j} M_{i}$. From the analysis of the $N_{j} $
maps, the following conclusions have been obtained: (i) 
vectors $\vec{n}_{2}$ and $\vec{n}_{3}$ are aligned in the direction
(0,0,1) for 1314 of these maps, which appear to have 
rather planar octopoles, (ii) in the remaining 207 cases, there
are no alignments and the octopole is less planar. In Fig.~\ref{f4},
one of these cases is displayed, the spots of the bottom panel are not aligned 
in the equatorial zone ($t=0.46$) and, then, the direction $\vec{n}_{3}$ is not
parallel to (0,0,1). Indeed, 
it has been numerically verified that these directions are almost orthogonal
to (0,0,1) in most of the above $207$ cases.
A theoretical proof of this orthogonality is not easy as a result 
of the particular form
of the {\em nonlinear} definition of $\vec{n}_{2}$, $\vec{n}_{3}$,
and $t$. 
In \S ~\ref{sec5}, 
this type of vector modes ($w$-modes) will be superimposed 
to simulate differential rotations and other 
symmetric divergenceless motions.
Then, the fraction of the superimpositions leading to 
$\vec{n}_{2}$ and $\vec{n}_{3}$ alignments will be 
experimentally found.

\section{3D superimpositions of vector modes}
\label{sec4}

According to Eq. (\ref{int33}),
functions $F_{pq}(\vec{r})$ can be calculated by using the
3D Fast Fourier Transform (FFT). In order to do
that, $512^{3} $ cells are considered inside a big box with a size
of $2 \times 10^{5} \ Mpc$. In this way, the cell size is $\sim
390 \ Mpc$ and, consequently, vector modes with spatial scales
between $10^{4} \ Mpc$ and $5 \times 10^{4} \ Mpc$ can be well
described in the simulation. We can then calculate function  
$F(\vec{r})$ to perform the integral in Eq. (\ref{int3}); 
in order to do that,
the observer is placed at an
arbitrary point located in the central part of the simulation box,
where the Fourier transform is expected to be well calculated and, then,
the integration is performed for each of the $3072$ 
directions of the pixel centers. 
The variations of
$F(\vec{r}) $ along the photon trajectories are smooth and,
consequently, the integrations giving $\Delta T / T$
can be easily performed. Furthermore, in a central cube with $
1.2 \times 10^{5} \ Mpc $ per edge ($60$ \% of the box size in our
simulations), we can place $5^{3} $ observers uniformly
distributed and separated by a distance of $3 \times 10^{4} \ Mpc $.
Then, quantity $\Delta T / T$ can be calculated for each of these
observers; thus, from a given simulation, the information we
obtain is greater than in the case of one unique observer located,
e.g., at the box center.

In this section, 
it is assumed (as in paper I) that
$v^{\pm}_{c0R}$ and $v^{\pm}_{c0I}$ are four statistically
independent Gaussian variables with vanishing mean, and also that each of these
numbers has the same power spectrum. The form of this common
spectrum is $P(k)=Ak^{n_{v}}$, where $n_{v} $ is the spectral
index of the vector modes and $A$ is a normalization constant. 
Two values of the spectral index:
$n_{v} =1 $ and $n_{v} =2 $ have been considered.
The spatial scales are varied from 
$10^{4} \ Mpc$ to $5 \times 10^{4} \ Mpc$ in all cases
(only very small wavenumbers are considered).
Four realizations of this 3D random superimposition of vector modes
have been performed for each spectrum and, then, 125 observers 
have been located as described above in each of the 
simulation boxes. Thus, $500$ simulations of the CMB 
relative temperature variations obtained from the 
last term of Eq. (\ref{dtt}) have been obtained. 
Moreover, the corresponding $500$ simulations of the term 
$- \vec {v}_{ce} \cdot \vec{n}$ have been also found.
In all cases, 
linearity conditions $|\vec{h}(\vec{r})| <<< 1$
and $|\vec{v}_{c}(\vec{r})| <<< 1$) have been verified using the 
relations:
\begin{equation}\label{vech2}
\vec {h}(\eta, \vec{r})= - 6 H_0^2 \Omega_{m} \, a^{-2}(\eta) \,
\int \frac{v_{c 0}^\pm(\vec{k})}{k^2} \, \vec{\epsilon}^{\, \pm}
(\vec{\kappa}) \exp (i \vec{k} \cdot \vec{r}) \, d^3 k  
\end{equation}
and
\begin{equation}\label{invel}
\vec{v}^{\pm}_{c}(\eta,\vec{r})= a^{-1} (\eta) \int v_{c0}^{\pm}
(\vec{k}) \, \vec{\epsilon}^{\, \pm}(\vec{\kappa}) \exp (i \vec{k}
\cdot \vec{r}) \, d^3 k  .
\end{equation}                                           
The analysis of all these simulations have let to the 
following main results: (i) the term  $- \vec {v}_{ce} \cdot \vec{n}$
is negligible against the last term of Eq. (\ref{dtt}). In Fig.~\ref{f5},
we present one simulation of each of these terms for $n_{v} =1 $. 
Numbers in the 
bottom panel ($- \vec {v}_{ce} \cdot \vec{n}$ term) are much smaller 
than those of the top panel [last term of Eq. (\ref{dtt})]. Obviously,
this comparison is independent on the spectrum normalization.
We have verified that the average $\langle C_{2} \rangle$ 
corresponding to the $500$ maps of the term $- \vec {v}_{ce} \cdot \vec{n}$
is $\sim 1/600 $ times smaller than the average calculated from
Eqs. (\ref{int3})--(\ref{int33}); therefore, 
the term $- \vec {v}_{ce} \cdot \vec{n}$ is hereafter neglected
and our study is restricted to the maps obtained from 
Eqs. (\ref{int3})--(\ref{int33});
(ii) the angle 
$\alpha_{23}$ subtended by directions
$\vec{n}_{2}$ and $\vec{n}_{3}$ is smaller than 
$10^{\circ}$ in nine of the $500$ simulations for 
both spectral indexes: $n_{v} =1 $ and $n_{v} =2 $. 
These numbers are compatible with the $8.33$ cases 
expected for a random distribution of direction $\vec{n}_{3}$ 
around a fixed $\vec{n}_{2}$ (see \citet{oli04});
(iii) parameter $t$ appears to be greater than $0.94$
in $40$ and $42$ simulations in the cases 
$n_{v} =1 $ and $n_{v} =2 $, respectively.
These numbers are to be compared with $35$, which is the corresponding 
number obtained by \citet{oli04} in the case of 
an isotropic Gaussian random field.
All these considerations are independent on the 
normalization of the spectra.   

We can conclude that 3D random superimpositions of large 
scale vector models do not explain either the observed
alignment of $C_{2}$ and $C_{3}$ ($\alpha_{23} \simeq 10^{\circ}$)
or the unusually planar octopole ($t \simeq 0.94$). However,
the study of some 2D distributions of modes is worthwhile.

\section{2D superimpositions of vector modes}
\label{sec5}

Special superimposition 
of vector modes are now considered. They are 2D superimpositions  
leading to divergenceless motions in long sized zones, 
which are hereafter called {\em parallel vorticity
regions} (PVRs). In each of these regions there is a privileged direction. 
Inside the 
region, the angular velocity (describing the local vorticity there) is
parallel to the privileged direction everywhere. 
The $x_{3} $-axis (hereafter $z$-axis)
can be chosen to be parallel to the privileged direction.
Finally, the PVRs are assumed to be uniform along this axis in the
sense that all the orthogonal planes are equivalent. In short, 
inside the PVRs, the components of the angular velocity 
are $W_{x}=0$, $W_{y}=0$ and $W_{z}=W_{z}(x^{1},x^{2})$. 
This configuration appears if functions $v_{c0}^{\pm} (\vec{k})$
are chosen as follows:
\begin{equation}
v_{c0}^{+} (\vec{k}) = v_{c0}^{-} (\vec{k}) = v_{c0} (\vec{k}) \, 
\delta(\theta_{k} - \frac {\pi}{2}) \ .
\label{ktun}
\end{equation}
In this equation, angle $\theta_{k} $ is one of the spherical coordinates in momentum space 
($k$ and $\phi_{k}$ being the other two) and $\delta $ stands for the Dirac distribution.
By substituting the distributions in Eq. (\ref{ktun}) into Eqs. (\ref{w1})--(\ref{w3}),
the following relations are obtained in position space:
\begin{equation}
W_{x} = W_{y} = 0
\label{wxy2}  
\end{equation}
\begin{equation}
W_{z} (x_{1},x_{2}) = \sqrt{2} \int v_{c0}(k_{1},k_{2},0) \, \sigma^{2} e^{i(k_{1}x_{1}+
k_{2}x_{2})} \, dk_{1} \, dk_{2} \ .   
\label{wztun}
\end{equation}
Analogously, from Eqs. (\ref{ktun}) and (\ref{int33}), the non-vanishing components of
$F_{pq}$ appear to be:
\begin{equation}
F_{11}=-F_{22}=-\sqrt{2} \int \frac {k_{1} k_{2}}{\sigma^{2}} \,
v_{c0}(k_{1},k_{2},0) \, e^{i(k_{1}x_{1}+
k_{2}x_{2})} \, dk_{1} \, dk_{2} \ ,
\label{F11} 
\end{equation}
\begin{equation}
F_{12}=\sqrt{2} \int \frac {k_{1}^{2}}{\sigma^{2}} \,
v_{c0}(k_{1},k_{2},0) \, e^{i(k_{1}x_{1}+
k_{2}x_{2})} \, dk_{1} \, dk_{2} \ ,  
\label{F12} 
\end{equation}  
\begin{equation}
F_{21}=-\sqrt{2} \int \frac {k_{2}^{2}}{\sigma^{2}} \,
v_{c0}(k_{1},k_{2},0) \, e^{i(k_{1}x_{1}+
k_{2}x_{2})} \, dk_{1} \, dk_{2} \ .
\label{F21} 
\end{equation}
Finally, vector $\vec{G}(\vec{r})$ involved in Eqs. (\ref{int6})--(\ref{int66})
has the following components:
\begin{equation}
G_{x} = G_{y} = 0
\label{gxy}  
\end{equation} 
\begin{equation}
G_{z} (x_{1},x_{2}) = - \sqrt{2} \int v_{c0}(k_{1},k_{2},0) \, e^{i(k_{1}x_{1}+
k_{2}x_{2})} \, dk_{1} \, dk_{2} \ .   
\label{gz3} 
\end{equation}    

As it follows from Eq.~(\ref{ktun}), our 2D superimpositions are 
combinations of the $w$-modes studied at the 
end of \S ~\ref{sec3} ($k_{3}=0$ and $v_{c0}^{+}  = v_{c0}^{-} = v_{c0}$)
and, consequently, vectors $\vec{n}_{2} $ and $\vec{n}_{3} $ are expected 
to be either parallel or orthogonal (almost in all cases). 
The proportions between alignments and
no alignments will be numerically obtained from the analysis of 
simulations.

\subsection{Differential rotations}
\label{sec51}

A present  
angular velocity of the form $W_z = W_{z}(\rho)$ is assumed, where
$\rho  = \sqrt{x_{1}^{2}+x_{2}^{2}}$.
This velocity describes a particular PVR, 
which could be interpreted
as a big region undergoing a differential rotation.
The local vorticity only depends on the distance to the
$z$-axis, which plays the role of 
the rotation axis. Then, from Eq. (\ref{wztun}) one easily 
finds
\begin{equation}
v_{c0}(k_{1},k_{2},0)=\frac {\sqrt{2}} {8\, \pi^{2} \, \sigma^{2} } 
\int W_{z}(\rho) \, e^{-i(k_{1}x_{1}+
k_{2}x_{2})} \, dx_{1} \, dx_{2} \ .
\label{vcfwz}  
\end{equation} 
Function $v_{c0}(k_{1},k_{2},0)$ is calculated 
by using the last equation and,
then, this function is substituted into Eqs.~(\ref{F11})--(\ref{F21}),
to get the $F_{pq}$ components. It is also substituted into Eq.~(\ref{gz3}) 
to obtain $G_{z}$. All these functions only depend on 
the coordinates $x_{1}$ and $x_{2} $. They are easily extended 
inside a 3D cube (where photons move) taking into account that the
planes orthogonal to the $z $-axis are indistinguishable.
For example, in the case of function $F_{12}$, its value  
at any 
point with coordinates $(x^{1},x^{2},x^{3})$ located inside the
3D cube would be 
$F_{12}(x^{1},x^{2},x^{3})=F_{12}(x^{1},x^{2},0)$.  
These extended functions allow us to calculate 
either $\Delta T / T $ (from Eq. (\ref{int3})) or
the polarization rotation angle $\delta \psi$ (from Eq.~(\ref{int6})).
These calculations can be performed for any observer located well 
inside the cube; in other words, for any observer whose last scattering
surface is fully localized inside the cube.

It is worthwhile to notice that, in the case of the rigid rotation of 
a big region,
the angular velocity $W_{z} $ vanish in a certain gauge,
in which the observer rotates with the region. In this gauge,  
Eq. (\ref{vcfwz}) gives: $v_{c}(k_{1},k_{2},0)= 0$ and, 
taking into account that this quantity 
is gauge invariant, it vanishes in any gauge; therefore,
according to Eqs.~(\ref{F11})--(\ref{gz3}) plus Eqs. (\ref{int3}) and (\ref{int6}), 
quantities $\Delta T / T$ and $\delta \psi$ vanish. In short, there is no either CMB anisotropy 
or Skrotskii rotations associated to rigid rotations (the same is valid
for rotations of the spatial coordinates in the absence of vector modes). 
These effects only appear in the
case of differential rotations, which cannot be globally avoided 
by any rotation of the reference frame.

Two functions $W_z = W_{z}(\rho)$ have been used: the first one is
\begin{equation}
\label{pro1}
W^{^{N}}_z (\rho) = \cases{N_{1} \Big[ e^{-(\rho^{2}/2m^{2})} - e^{-2} \Big]   
& \,\,\,\,\,  $\rho \le 2m$ \cr 0 & \,\,\,\,\, $ \rho > 2m$
\cr}  \ ,
\end{equation}   
where $N_{1}$ is a normalization constant. The length $m$ defines the 
spatial size of the PVR. The values $m=5 \times 10^{3} \ Mpc $ (case NI) and 
$m=3 \times 10^{3} \ Mpc $ (case NII) have been tried. Evidently, the spatial 
scales involved in this differential rotation are very large. 
The second function is:
\begin{equation}
\label{pro2}
W^{^{C}}_z (\rho) = \cases{N_{2} \cos { (\pi \rho / 2 \rho_{max}) }  
& \,\,\,\,  $\rho \le \rho_{max}$ \cr 0 & \,\,\,\, $ \rho > \rho_{max}$
\cr} \ ;
\end{equation}   
quantities $N_{2}$ and $\rho_{max}$ being the normalization constant
and the parameter defining the spatial profile of the angular velocity,
respectively. Two values of $\rho_{max}$ have been studied: 
$\rho_{max} = 6.8 \times 10^{3} \ Mpc $ (case CI) and 
$\rho_{max} = 4 \times 10^{3} \ Mpc $ (case CII).

Once an angular velocity profile $W_z = W_{z}(\rho)$ has been assumed
(cases NI, NII, CI, and CII),
only two elements remain free: (i) the normalization constant, and (ii)
the location of the observer in the simulation square. The square is that 
appropriate for the Fourier transforms in Eqs.~(\ref{F11})--(\ref{gz3}). 
For the above 
profiles, a square size of $5 \times 10^{4} \ Mpc $ is used and, then,
81 observers are uniformly located in a central square of 
$2 \times 10^{4} \ Mpc $ size. The separation between neighboring
observers is $2.5 \times 10^{3} \ Mpc$; therefore, 
once parameter $m$ ($\rho_{max}$) is fixed in the profile $W^{^{N}}_z$
($W^{^{C}}_z$),  
$81$ simulations of $\Delta T/T$ and $\delta \psi $ can be obtained as 
it has been described in the first paragraph of \S ~\ref{sec51}. 
Each map corresponds to a localization of the observer characterized by its
distance to the rotation axis ($x_{1}=x_{2}=0$ line). 
The analysis of the resulting HEALPIx maps has let to the following 
main conclusions:
(1) the $C_{2}$-$C_{3}$ alignment is perfect for any of the above 
$W_{z} $ profiles and observers ($\alpha_{23} = 0$),
(2) the inequality $t>0.94 $ also is satisfied in all cases. 
These results are encouraging. The proposed differential rotations 
plus appropriate large scale scalar modes could easily 
lead to the observed angle $\alpha_{23} \simeq 10^{\circ}$ and also
to the parameter $t \simeq 0.94 $. Of course, 
the large scale vector modes under consideration should dominate 
against the scalar ones. Thus, the alignment produced by 
the differential rotation (vector modes) would not be hidden by 
the effects of standard scalar modes.
The amplitude of the scalar perturbations contributing to small 
$\ell $ multipoles (very large scales) should be smaller than
those corresponding to the standard flat spectrum (compatible with 
the remaining observed $C_{\ell} $ quantities). 
Either a certain cutoff or a
damping of the scalar fluctuations would be necessary on very large scales.
Details about the possible cutoff scale or the gradual damping
are out of the scope of this paper; however,
the general considerations of this paragraph are important to normalize the 
$W_z $ profiles.

A few considerations about recent CMB observations are 
necessary before describing our normalization method. 
According to \citet{hin06} (WMAP three years data
analysis), the CMB quadrupole is $C^{^{WMAP}}_{2} \simeq 2.96 \times 10^{-11}$
whereas the octopole is $C^{^{WMAP}}_{3} \simeq 7.38 \times 10^{-11}$;
hence, if it is assumed that the contribution
of scalar and vector modes to these multipoles are to be added (statistical independence of the 
scalar modes and the differential rotation) and, moreover, it is taken into account
that the contribution of the vector modes must dominate (see previous paragraph), 
such a vector contribution should roughly satisfy the following conditions:
(a) $C_{2} $ must be a little smaller than $2.96 \times 10^{-11}$ and, (b)
$2C_{2} < C_{3} < 3C_{2}$; hence, the following 
method is used to normalize in each of the cases 
NI, NII, CI, and CII: in a first step, the $C_{2}$ and $C_{3}$ multipoles 
of the 81 maps are calculated for an arbitrary 
normalization and, then, the maps (observers) compatible with condition 
(b) --which is independent on normalization-- are found. 
The total number, $N_{b}$, of these maps is given --for each case-- in Table \ref{tab1}.
Some of these maps correspond to observers located at the
same distance from the rotation axis and, consequently, 
their normalizations 
are identical except for small numerical errors. This fact has been verified.
The total number of distinct distances (observers), $N_{d} $, 
and the distances themselves, $d_{or}(i)$ with $i:1,N_{d}$, are 
also given in Table \ref{tab1}.
In a second step, the normalization constant is chosen to have 
$C_{2} = 2.5 \times 10^{-11}$ for each of the above $N_{d}$ observers
and, then, the resulting octopoles, $C_{3}(i) $,  are
calculated and shown in Table \ref{tab1} for $i:1,N_{d}$.
A number $N_{d} $ of different normalizations is thus obtained.
Each of these normalizations is separately considered. The 
$\Delta T/T$ and $\delta \psi $ maps corresponding to one of 
the two observers of case NI ($i=1$ in Table \ref{tab1})
are displayed in Fig.~\ref{f6}. Top panel shows a $\Delta T/T$
map which seems to be clearly compatible with a planar 
octopole (estimated value: $t \simeq 0.9979$) and a perfect alignment 
($\alpha_{23}=0$, with possible small errors due to
the limited angular resolution of the HEALPIx maps).
Bottom panel displays the corresponding $\delta \psi$ map. 
Angles close to $0.1$ degrees are reached in some directions,
the angles are similar (a little smaller) than those obtained 
in paper I, which were estimated by using a rather arbitrary normalization. 

\clearpage

\begin{deluxetable}{ccccccccc}
\tabletypesize{\scriptsize}
\tablecaption{2D simulations based on $W_{z}$ profiles.\label{tab1}}
\tablewidth{0pt}
\tablehead{
\colhead{CASE} & \colhead{$N_{b}$\tablenotemark{\alpha}} & 
\colhead{$N_{d}$\tablenotemark{\beta}} 
& \colhead{$C_{3}(1)\times 10^{11}$} & \colhead{$C_{3}(2)\times 10^{11}$} &
\colhead{$d_{or}(1)\times 10^{-3}$} & \colhead{$d_{or}(2)\times 10^{-3}$} &
\colhead{$A_{wz}(1)\times 10^{9}$} & \colhead{$A_{wz}(2)\times 10^{9}$} 
}
\startdata
NI &12 & 2 &$7.23 $ & $5.94$  
 & $7.9 \ Mpc$ & $9.0 \ Mpc$ & $0.99$ & $0.95$\\
NII &8 & 1 &$6.58 $ &-- & $7.9\ Mpc $ & -- &$2.51$ & --\\
CI &12 & 2 &$7.04 $ &$5.33 $ &$7.9 \ Mpc
$ &$ 9.0 \ Mpc$ & $1.51$ & $1.41 $\\
CII &8 & 1 &$5.77 $ & -- &$7.9 \ Mpc$ & -- &$3.94$ & --\\   
\enddata
\tablecomments{First column lists the four $W_{z}$ profiles defined
in the text. In each case, 81 observers are uniformly distributed in the 
central part of the simulation box}
\tablenotetext{\alpha}{The number of observers whose CMB 
multipoles satisfy
the relation $2C_{2} < C_{3} < 3C_{2}$ is $N_{b}$}
\tablenotetext{\beta}{Among the $N_{b}$ observers, there are $N_{d}$ ones 
which are actually
different (they are located at distinct distances, $d_{or}$, from the rotation
axis)} 
\tablecomments{$C_{3}(1)$ is the octopole (after normalization by the condition
$C_{2}=2.5\times 10^{-11}$) of one of the $N_{d}$ observers, whereas $C_{3}(2)$
corresponds to the second of these observers (if it exists). The same for
$d_{or}$ and for the dimensionless ratio $W_{z}(\rho=0)/H_{0}$}

\end{deluxetable}    

\clearpage

After the above normalization method has been applied, any of the $N_{d}$ 
normalizations corresponds to an observer (characterized by its distance 
to the rotation center) whose $C_{2} $ and $C_{3} $ multipoles 
satisfy the following conditions: (i)
they are appropriate to explain 
the values observed by the WMAP satellite with the help of a certain 
contribution due to scalar modes (to be estimated),
(ii) these multipoles are fully aligned, and (iii) 
the octopole is very planar ($t>0.94$).
The distances from the observers to the rotation axis are different from 
zero (see Table \ref{tab1}) and, consequently,
these observers are not placed on the rotation axis but in another 
position, which is so much probable as any other position in the space.

Normalizations lead to the values of the constants $N_{1} $ and
$N_{2} $ involved in Eqs.~(\ref{pro1})--(\ref{pro2}), from which, 
the dimensionless amplitude of the angular velocity profile 
$A_{wz} = W_{z}(\rho=0)/H_{0} $ can be found in each case.
The resulting $A_{wz} $ values are given in Table~\ref{tab1}
for the normalizations included in it.
They are a few times greater than the value $4.3 \times 10^{-10}$ reported by
\citep{jaf05} in the framework of a fully different model.

\subsection{Statistical parallel vorticity fields}
\label{sec52}

In this section, a PVR region is simulated by using statistical methods.
The components $v_{c0R}$ and $v_{c0I}$ of 
the complex numbers $v_{c0}(k_{1},k_{2},0)$
are generated as two statistically
independent Gaussian variables with the 
same power spectrum and zero mean. The form of the spectrum is the same as in the 
3D simulations; namely,
$P(\sigma)=A\sigma^{n_{v}}$, and the chosen spectral indexes and 
spatial scales are also the same as in the 3D statistical realizations.

Ten realizations of these 2D random superimposition of vector modes
have been performed for each spectrum ($n_{v}=1$ and $n_{v}=2$) and, then, 
81 observers 
have been uniformly located in the simulation square using the 
same method as in the 2D simulations with 
$W_{z} $ profiles; however, the sizes of the simulation square 
and the central square
are $2 \times 10^{5} \ Mpc $ and $1.28 \times 10^{5} \ Mpc $, respectively,
and the distance between observers is 
$1.6 \times 10^{3} \ Mpc$. 
Thus, $810$ simulations of the CMB 
relative temperature variations produced by PVRs have been 
obtained. The corresponding $\delta \psi$ maps have been also found.
All these maps have been analyzed. 
Results from this analysis are now described; we begin with
various conclusions which are independent on the spectrum normalizations: 
($\alpha$) the angle $\alpha_{23} $ is zero
in the $48.64 \%$ ($48.4 \% $) of
the $810$ simulations for 
$n_{v} =1 $ ($n_{v} =2 $),
($\beta$) parameter $t$ appears to be greater than $0.94$
in the $18.64 \%$ ($19.88 \%$) of the simulations for
$n_{v} =1 $ ($n_{v} =2 $), and 
($\gamma $) the conditions 
$t>0.94$ and $\alpha_{23} =0$ are simultaneously satisfied
in the $\sim 11 \% $ ($\sim 12 \% $) of the simulations for
$n_{v} =1 $ ($n_{v} =2 $). These last percentages 
can be found from Table \ref{tab2}, where the number
of cases, $n_{at}$, satisfying the two relations $t>0.94$ and $\alpha_{23} =0$ 
is given for each of the ten 2D realizations. We have counted 
these cases because, as it has been discussed in \S ~\ref{sec3}, 
conditions $t>0.94$ and $\alpha_{23} =0$ do not seem to be 
independent and, consequently, the probability 
of the realizations satisfying the two relations is 
not {\em a priori} the product of the individual probabilities.

The spectra are normalized as follows: first,
all the simulations satisfying the conditions
$t>0.94$ and $\alpha_{23} =0$ are normalized 
by the condition $C_{2}=2.5\times 10^{-11}$ and, then, those of them 
satisfying the inequalities $2C_{2} < C_{3} < 3C_{2}$ 
are identified and counted. Their total number, $n_{obs}$, is 
given in Table \ref{tab2} for 
each of our ten 2D realization. It is worthwhile to notice that,
each of the normalized simulations 
corresponds to one of the ten 2D statistical 
realizations and also to an observer located at
a certain position in the simulation cube. Since there is no a rotation 
axis, coordinates $x^{1}$ and $x^{2}$ are both necessary to
fix the observer position in the plane orthogonal to 
the vorticity direction of the PVR.

For $n_{v}=1$ ($n_{v}=2$), number $n_{obs}$ appears to be zero in five (one) 
of our ten 2D statistical superimpositions of vector modes. In these five (one) 
cases, conditions $t>0.94$ and $\alpha_{23} =0$ are satisfied (see Table \ref{tab2}),
but there are no observers measuring a quadrupole 
$C_{2}=2.5\times 10^{-11}$ and an octopole
satisfying the relations $2C_{2} < C_{3} < 3C_{2}$. 
It is then easily calculated the probability of having at least an 
observer whose measurements satisfy
the four conditions $t>0.94$, $\alpha_{23} =0$,
$C_{2}=2.5\times 10^{-11}$ and, $2C_{2} < C_{3} < 3C_{2}$,
namely, whose measurements may be 
compatible with current observations after introducing 
appropriate sub-dominant scalar modes. This probability 
is close to $\sim 5.5 \%$  ($\sim 10.8 \%$) for $n_{v}=1$ ($n_{v}=2$).
With these probabilities we cannot say that we live in a 
very special zone of the PVR, but in a reasonably 
probable one, which is equally probable than any other 
positions inside the PVR.

For $n_{v}=2$ and the 2D realization number $9$ of Table \ref{tab2},
there are two observers ($n_{obs}=2$) whose measurements are compatible with
the four above conditions. One of these observers, located at 
$\sim 5 \times 10^{4} \ Mpc $ from the cube center, would measure
$t \simeq 0.9631$, $\alpha_{23} =0$, $C_{2}=2.5\times 10^{-11}$, and 
$C_{3} = 6.45 \times 10^{-11}$. The 
$\Delta T/T$ and $\delta \psi $ maps corresponding to this 
observer are shown in Fig.~\ref{f7}. Top panel displays 
the $\Delta T/T$
map, which looks like those compatible with a planar 
octopole and a perfect alignment.
The corresponding $\delta \psi$ map is exhibited in the bottom panel. 
The largest angles -- close to $\sim 4.4 \times 10^{-3}$ degrees--
are much smaller (by a factor $\sim 1/50$) than those based on the 
normalization of paper I. Of course, these angles are too small 
to produce any currently significant $B$-polarization of the CMB. 

\clearpage

\begin{deluxetable}{cccccc}
\tabletypesize{\scriptsize}
\tablecaption{Statistical 2D simulations.\label{tab2}}
\tablewidth{0pt}
\tablehead{
\colhead{CASES ($n_{v}=1$)} & \colhead{$n_{at}$\tablenotemark{\alpha}} 
& \colhead{$n_{obs}$\tablenotemark{\beta}} 
& \colhead{CASES ($n_{v}=2$)} & \colhead{$n_{at}$} &
\colhead{$n_{obs}$}
}
\startdata
1 &7 & 1 &1 &11 &3\\
2 &9 & 0 &2 &10 &1\\
3 &6 & 2 &3 &12 &2\\
4 &9 & 1 &4 &12 &2\\  
5 &9 & 0 &5 &11 &2\\
6 &7 & 3 &6 &10 &1\\
7 &9 & 0 &7 &12 &2\\
8 &4 & 0 &8 &12 &1\\ 
9 &11 & 1 &9 &12 &2\\
10 &17 & 0 &10 &12 &0\\
\enddata
\tablecomments{Ten 2D statistical 
simulations corresponding to the spectral indexes $n_{v}=1$ and $n_{v}=2$
are numbered in the 
first and fourth columns, respectively}
\tablenotetext{\alpha}{Number of observers which measure $\alpha_{23} = 0$ and
$t>0.94$ (among 81 of them placed inside the simulation box)}
\tablenotetext{\beta}{Number of cases (among 81), in which measurements 
would be compatible with conditions $\alpha_{23} = 0$,
$t>0.94$, and $2C_{2} < C_{3} < 3C_{2}$ } 
\end{deluxetable}    

\clearpage

Finally, Fig.~\ref{f8} shows a dimensionless quantity proportional 
to the present angular velocity $W_{z} $. The represented zone 
is located inside the simulation square and centered in it.
The normalization is the same as in Fig.~\ref{f7} (same 
2D simulation and observer).
Red and blue spots correspond to regions which 
are rotating in opposite senses. A boundary with $W_{z}=0 $
separates them. The mean value of $(W_{z}/H)_{0}$ is
negligible by construction and the typical deviation is
$\langle |W_{z}|^{2} \rangle^{1/2}/H_{0} = 3. \times 10^{-9} $.
Many realizations (as that of the Figure) 
have been considered to conclude that
the typical value of $(W_{z}/H)_{0} $ is always a few
times $10^{-9}  $.

\section{Discussion and conclusions}
\label{sec6}

Appropriate combinations of large scale vector perturbations 
have been introduced in the concordance model and, then, 
their effects on the CMB anisotropy have been studied in detail.
Our main conclusions can be summarized as follows:
3D superimpositions of vector modes do not explain 
the CMB anomalies; however, some 2D superimpositions 
of these modes lead to good results. Two types of 2D simulations 
have been performed: one of them represents differential rotations 
of big regions and the other one leads to extended statistical
PVRs. In these two cases there is
a preferred direction of symmetry. It is the direction of the 
angular velocity, which is the same in any point of the 
perturbed region. In the first case, there is a 
symmetry around the rotation axis in the plane orthogonal 
to the preferred direction,
however, statistical  
PVRs do not introduce such a rotational symmetry. 

Suitable differential rotations can explain the planar character
of the octopole, its alignment with the quadrupole, and the main part 
of the $C_{2}$ and $C_{3}$ values 
observed with WMAP. These facts are proved, in 
\S ~\ref{sec51}, for two different $W_{z} $ profiles. Polarization
rotation angles $\delta \psi $ close to $0.1$ degrees 
are produced by these profiles. Other possible profiles could
produce slightly greater angles. A sub-dominant contribution 
of large scale scalar modes could then account for a small part
of the observed quadrupole and octopole, which would be complementary 
of the part due to vector modes. These scalar modes
could be also responsible for the observed angle 
$\alpha_{23} \simeq 10^{\circ} $, which vanishes for pure
differential rotations. The required scalar modes would 
destroy the rotational symmetry in the plane orthogonal 
to the axis of the differential rotation.
Skrotskii rotations close to $0.1$ degrees would produce a 
$B$-polarization of the CMB, which could be marginally observable 
by future satellites (see paper I).

For statistical PVRs, there is
an appreciable probability of accounting for all the anomalies
explained by differential rotations. This probability
depends on the form of the assumed power spectrum and also on
the interval of $k$ values considered in the computations. 
The dependence on the spectral index has been pointed out by considering
two distinct values $n_{v}=1$ and $n_{v} = 2$ (see \S ~\ref{sec52}). 
The mentioned probability 
is greater in the case $n_{v}=2 $ ($\sim 11 \%$). Of course, a certain level of 
scalar modes is necessary (as in the case of differential rotations)
in order to explain the observed angle $\alpha_{23} \simeq 10^{\circ}$.
Statistical PVRs lead to $\delta \psi $ angles which are too small to produce
significant levels of $B$-polarization. 
Other intervals of spatial scales, and 
other power spectra could lead to higher probabilities for
the explanation of anomalies and, perhaps, to greater 
Skrotskii rotations. In a certain $k $ interval, the spectrum 
of vector modes could have any form (a power law is not 
required either by any theoretical prediction or by observational
evidences). In a finite interval, e.g., between $10^{4}$ and $5\times 10^{4} \ Mpc$,
the spectral index of a power spectrum is arbitrary, nevertheless,
only some spectral indexes are admissible, as $k$ tends to zero,
to avoid divergences in some integrals (e.g., that of Eq. (\ref{F11})).
                                     
In \citet{rac07}, it is stated that, at high confidence, 
there is no any rotational 
symmetry of the CMB in the plane orthogonal to the symmetry axis. 
This fact is compatible with differential rotations by two reasons:
(i) the mentioned rotational symmetry would be only observed from 
points placed on the rotation axis, whereas we are not located on this 
line with very high 
probability, and (ii) there may be either large scale 
sub-dominant scalar perturbations or deviations with respect to a perfect
differential rotation and, obviously, these perturbations and deviations 
could contribute to hide any rotational symmetry and also 
to explain the deviation from zero observed in 
the angle $\alpha_{23}$.

The asymmetry of the 
the North and South ecliptic hemispheres is also compatible with 
our 2D superimpositions of vector modes. We predict two 
equivalent hemispheres, nevertheless, they are not separated by the 
ecliptic plane, but by the plane orthogonal to the angular velocity.
Furthermore, in some slightly different scenarios, 
the equivalence of these two hemispheres 
could disappear. It occurs, e.g., if the
last scattering surface of the observer is partially  
outside the PVR, which is particularly probable for
PVRs which are not too extended in some direction.

Solar system alignments would be casual, as it seems natural
in any cosmological explanation of the observed anomalies.
See \citet{cho07}. All the theories of this type (including our proposal) 
would be ruled out by solutions of the CMB anomaly problem based on both,
the ordinary spectrum 
of scalar perturbations, and a non cosmological component
accounting for the observed statistical correlations with the local
geometry of the solar system; however, current observations and data analysis
have not unveiled any component of this type accounting for the 
CMB anomalies.

We have assumed very large spatial scales to alter only a few 
low-$\ell$ multipoles; nevertheless, only vector modes have been 
considered. Why large scale scalar modes have not been tried? 
The main reasons are now pointed out. For the chosen spatial scales,
combinations of modes should lead  
to very large almost-homogeneous regions. It occurs 
whatever the nature of the perturbations may be. In the case
of vector modes, the angular velocity will 
be almost-homogeneous in these regions and, 
consequently, it will have almost the 
same direction everywhere. These absolutely 
natural regions, which could have sizes comparable to that of the 
sphere bounded by the large scattering surface (for large enough 
spatial scales),  
are the PVRs we need to explain anomalies.
In the case of scalar perturbations,
the density contrast should be almost-constant in these large 
regions and, consequently, a cylindrical scalar inhomogeneity would be actually 
unlikely. Moreover,          
a flattened inhomogeneity does not seem likely as 
a result of the small scales required by the short thickness
of the structure,
which would be small enough to affect multipoles with 
too large $\ell$ values. Hence,
symmetry axis and  
preferred planes seem to be rather improbable in the case of
large scale scalar modes. 
Although these arguments are qualitative they strongly suggest the  
use of vector modes.

Let us finish this paper with a list of a few 
open problems which should be addressed in the near future:
(1) the origin and evolution laws of the vector modes 
(e.g. brane-worlds, strings, and so on)
deserve particular attention. Only a consistent theory on 
these subjects
could give answers to important questions as: in what a 
cosmological period (or periods) are generated the vector modes? 
How do they actually decay? 
How much probable are the PVRs? Are 
scalar and vector modes statistically independent? 
(2) Multipole components for $\ell >3 $ must be also analyzed and 
compared with those extracted from WMAP data. Multipole 
vectors \citep{cop04} should be used in this extended study.  
(3) The proportions between large scale scalar and vector modes 
must be considered in more detail and,
(4) deviations from the perfect parallelism assumed in our 2D
superimpositions of vector modes could lead to interesting results
(hemisphere asymmetry, $\alpha_{23} $ observed value, and so on).

Large scale rotations are currently 
enigmatic (even for us), but the origin of the familiar cosmic expansion 
has kept unknown during a century. In both cases, 
rotations and expansion, rejection (acceptance) would be only justified 
by the disagreement (agreement) between predictions and observations
(without prejudices). Although we have not a closed theory on the 
subject of this paper, results related with the CMB anomalies are 
actually encouraging 
and, consequently, more study
is worthwhile.

\acknowledgments

This work has been supported by the Spanish Ministerio de
Educaci\'on y Ciencia, MEC-FEDER project FIS2006-06062.






\begin{figure}
\epsscale{.40}
\plotone{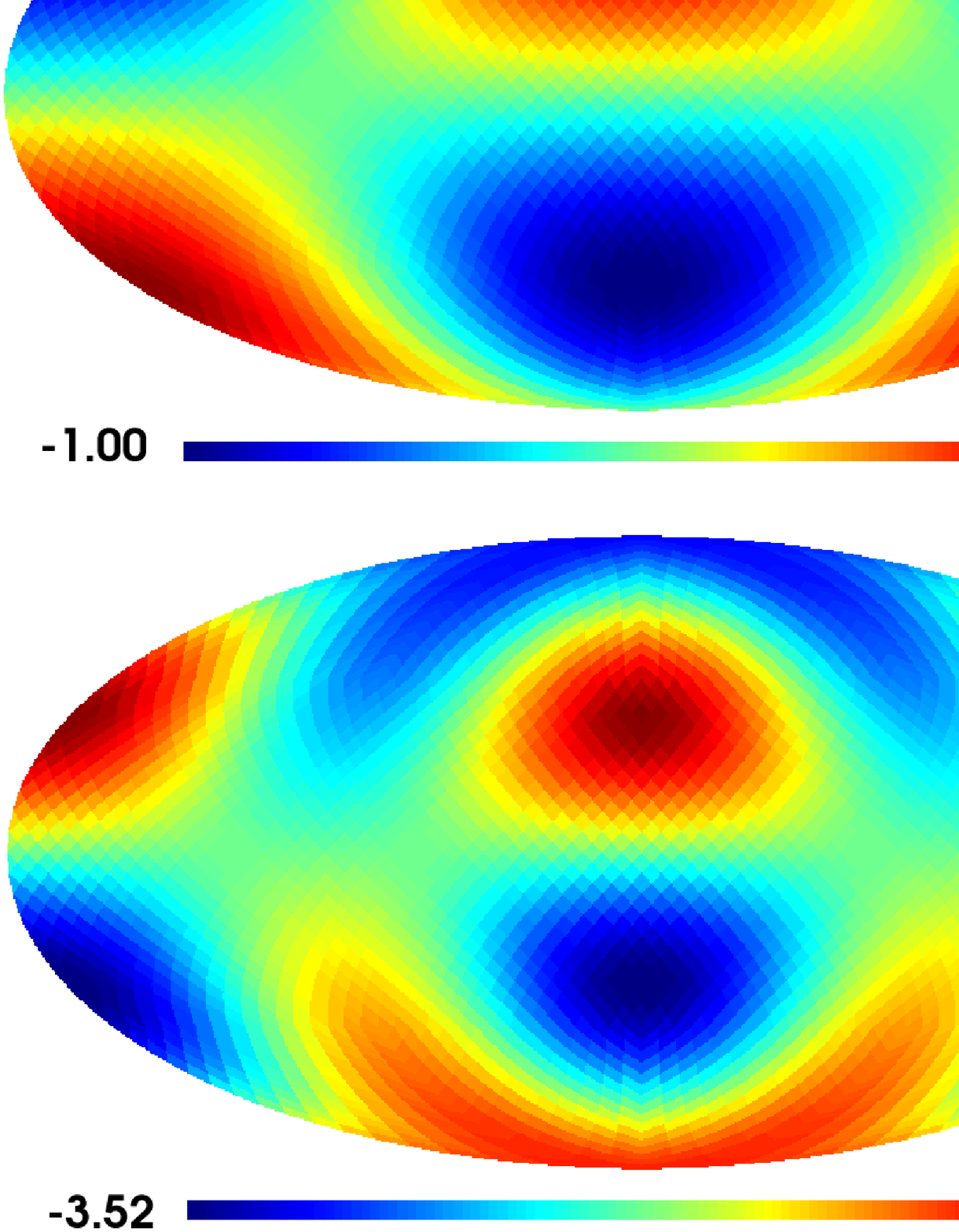}
\caption{Top panel shows the  HEALPIx map 
of $ (\Delta T/T) \times 10^{5} $ for the vector mode [1]
(see text). Middle (bottom) panel displays the quadrupole (octopole) of this map.
The alignment of $C_{2}$ and $C_{3}$ is evident. 
The octopole looks planar.
Normalization is irrelevant.
\label{f1}}
\end{figure}

\clearpage

\begin{figure}
\epsscale{.40}
\plotone{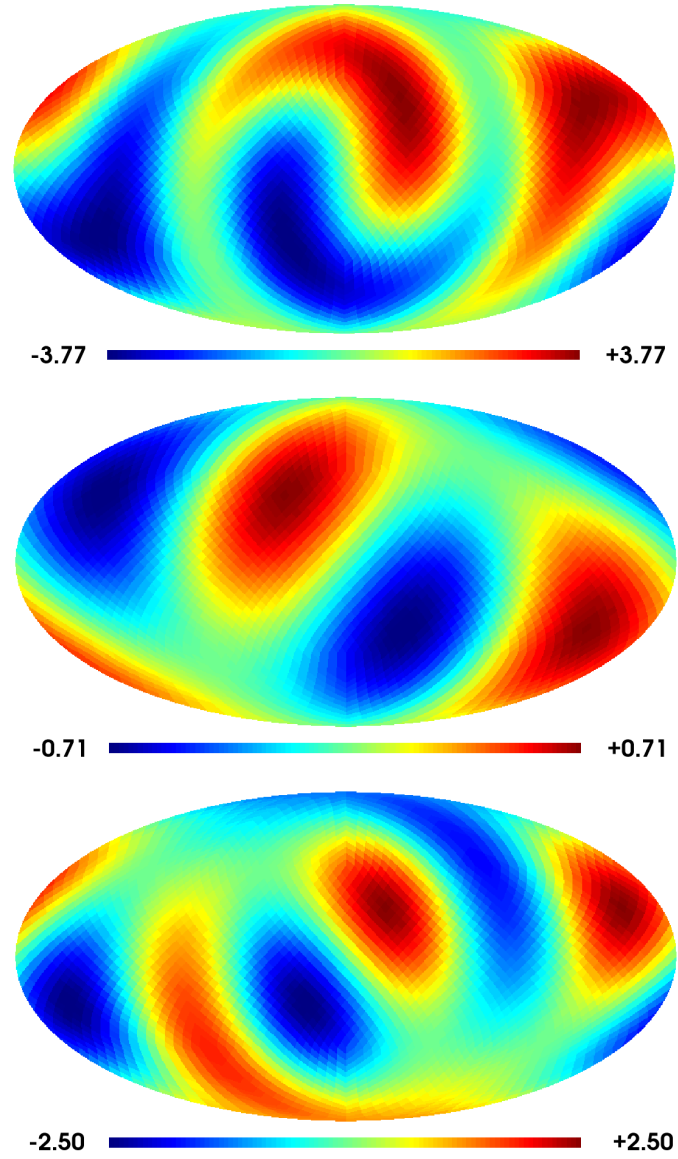}
\caption{Same as in Fig.~\ref{f1} for vector mode [2] (see text). 
There is no alignment in this case, but the octopole is 
visibly planar. Vectors $\vec{n}_{2} $ and $\vec{n}_{3} $ are 
almost orthogonal
\label{f2}}
\end{figure}
  
\clearpage  
  
\begin{figure}
\epsscale{.40}
\plotone{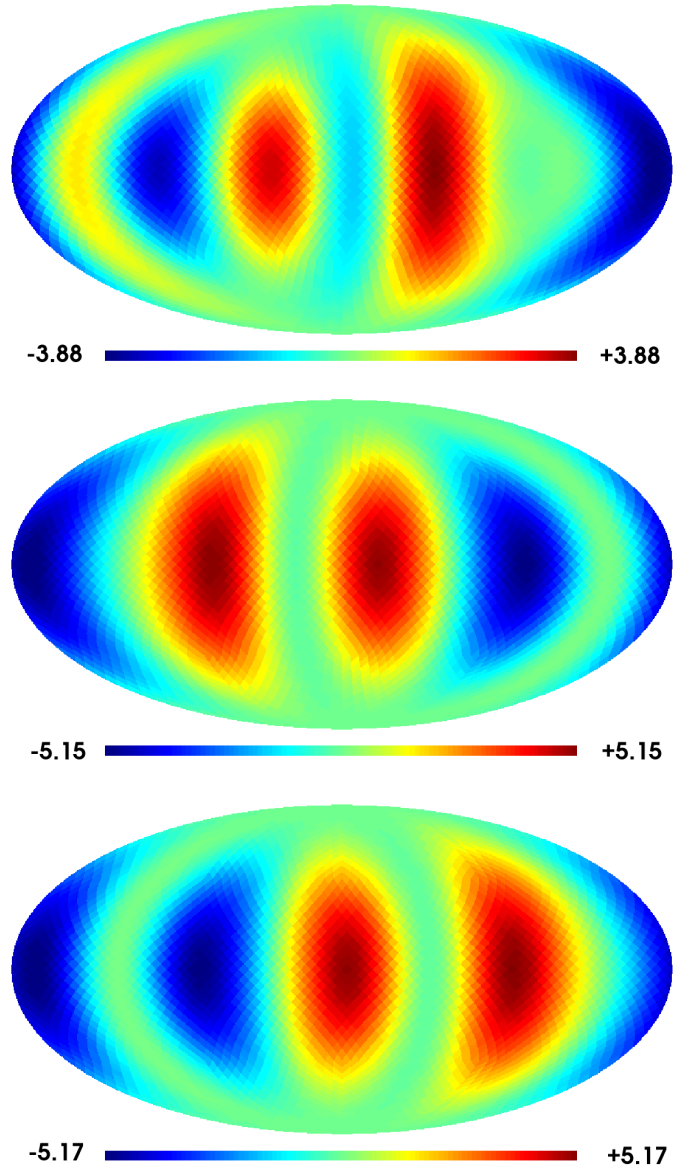}
\caption{Each panel shows the HEALPIx map of $(\Delta T/T)\times 10^{5}$
produced by a different $w$-mode (see text). The 
equatorial alignment and the planar character of the octopole are 
evident in the three panels.
Normalization is irrelevant.
\label{f3}}
\end{figure}

\clearpage

\begin{figure}
\epsscale{.40}
\plotone{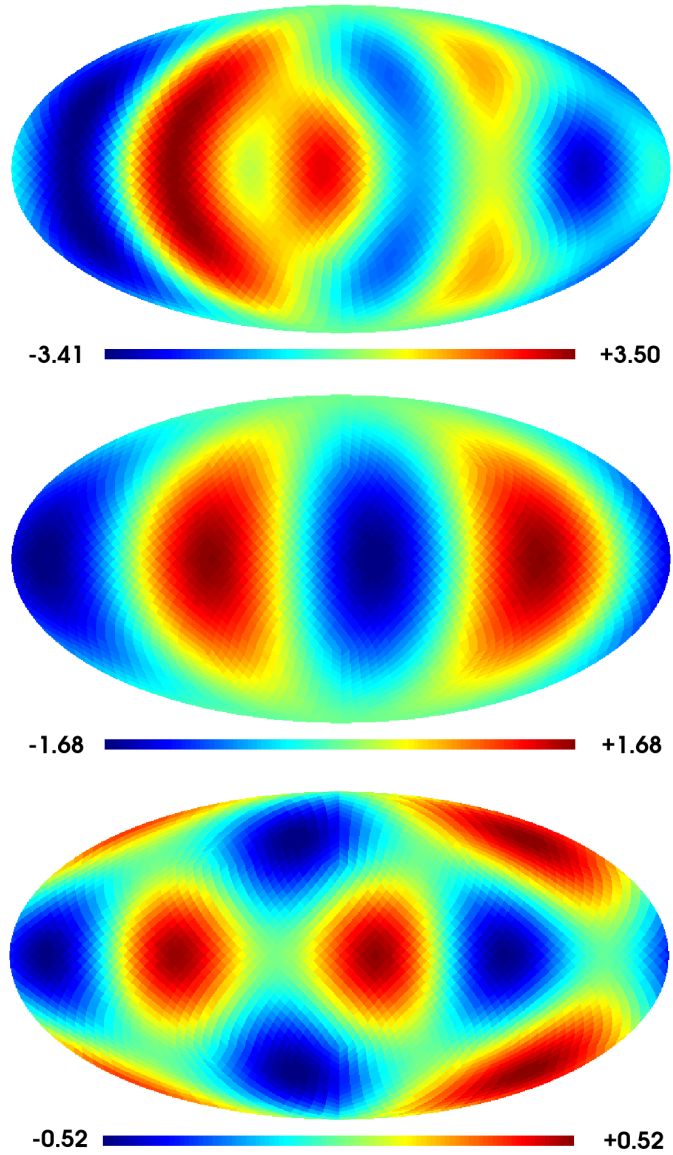}
\caption{Same as in Fig~\ref{f1} for the superimposition of $w$-modes $N_{1126}$
(see \S ~\ref{sec3}). Vector $\vec{n}_{2} $ has the direction (0,0,1) and 
the octopole is not planar. Calculations  
indicate that vectors $\vec{n}_{2} $ and $\vec{n}_{3} $ are orthogonal; hence,
direction $\vec{n}_{3} $ is contained in the equatorial plane.
Normalization is irrelevant.
\label{f4}}
\end{figure}

\clearpage

\begin{figure}
\epsscale{.40}
\plotone{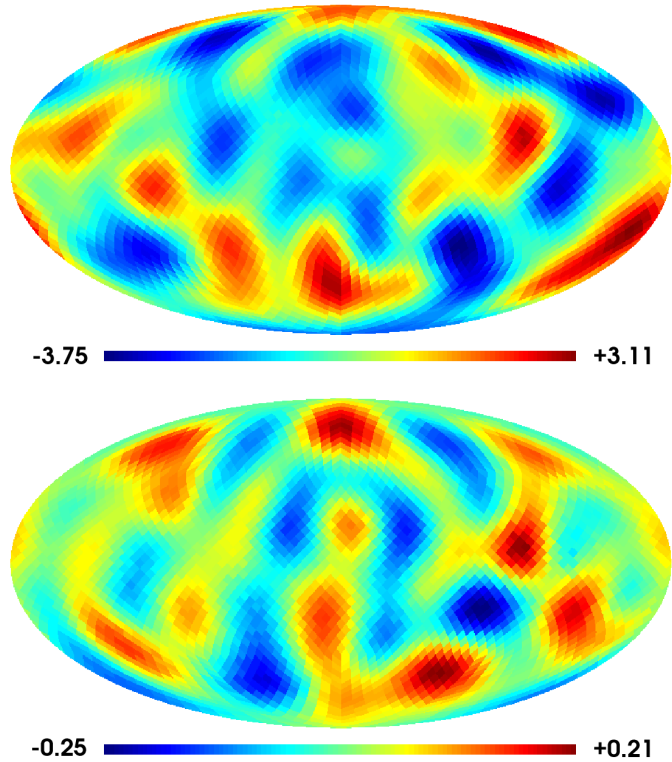}
\caption{Both panels are HEALPIx maps of 
quantity $ (\Delta T/T) \times 10^{5} $.
Top panel shows the part of this quantity
due to the last term of Eq.~(\ref{dtt}), whereas
the part obtained from the term $- \vec {v}_{ce} \cdot \vec{n}$
is displayed in the bottom panel. The second part 
is much smaller than the first one. That is independent 
on the power spectrum normalization.
\label{f5}}
\end{figure}

\clearpage

\begin{figure}
\epsscale{.40}
\plotone{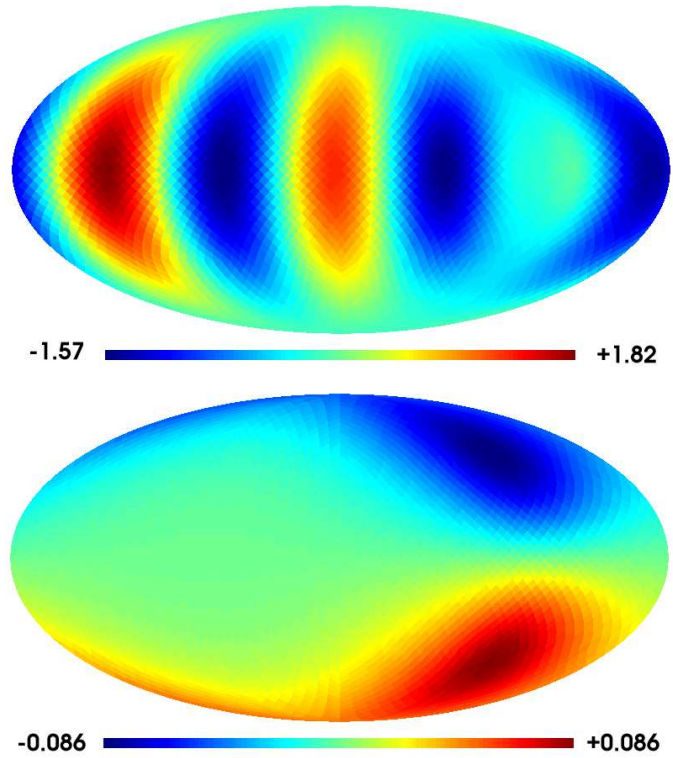}
\caption{Top panel is a HEALPIx map 
of $ (\Delta T/T) \times 10^{5} $ for
case NI and observer $i=1$ (see Table~\ref{tab1}).
A strong $C_{2}$--$C_{3}$ alignment and a high 
$t$ value are evident.
The corresponding $\delta \psi $ map is displayed
in the bottom panel, where the angles are given in
degrees.
\label{f6}}
\end{figure}

\clearpage

\begin{figure}
\epsscale{.40}
\plotone{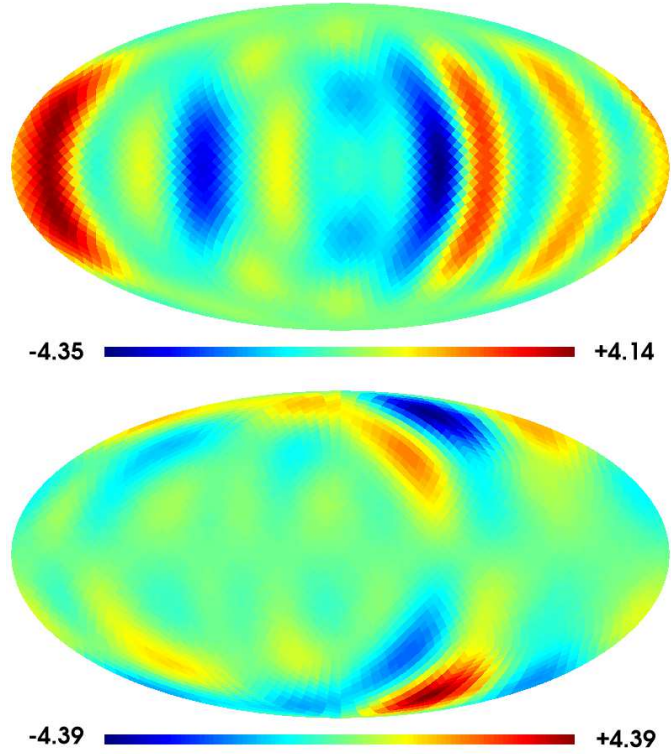}
\caption{Top: same as in Fig.~\ref{f6} for one of the two
observers measuring multipoles $C_{2}$ and $C_{3}$
compatible with current observations in the realization number
$9$ of the case $n_{v}=2$ (see Table~\ref{tab2} and text). 
Bottom: the corresponding Skrotskii angles, $\delta \psi $, 
are given in units of $10^{-3}$ degrees.
\label{f7}}
\end{figure}

\clearpage

\begin{figure}
\epsscale{.60}
\plotone{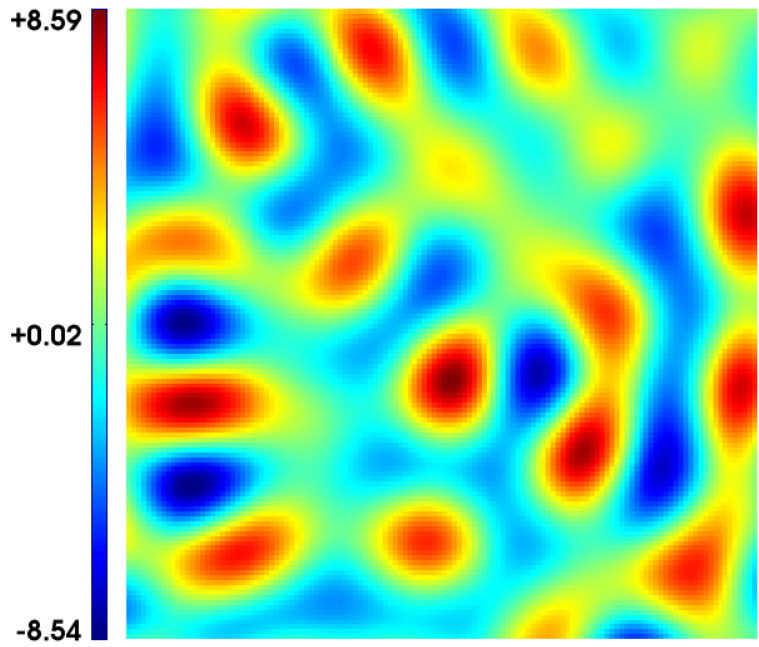}
\caption{Map of the dimensionless quantity $(W_{z}/H)_{0} \times 10^{9}$ 
in the plane orthogonal to the 
angular velocity.  The size
of the represented square is $50 \ Mpc$
\label{f8}}
\end{figure}

\end{document}